\documentclass[pre,twocolumn]{revtex4-2}
\usepackage{amsmath}
\usepackage{amssymb}
\usepackage{graphicx}
\usepackage{bm}
\usepackage{physics}
\usepackage{mathtools}
\usepackage{color}
\usepackage[caption=false]{subfig}

\usepackage{afterpage}

\newcommand{\llrr}[1]{\langle \! \langle #1 \rangle \! \rangle}

\usepackage{hyperref}
\definecolor{myblue}{RGB}{0,0,255}
\hypersetup{
    colorlinks,
    citecolor=myblue,
    linkcolor=myblue,
    urlcolor=myblue
}

\begin{document}

\title{Accelerated Jarzynski estimator with deterministic virtual trajectories}
\author{Nobumasa Ishida}
\email{ishida@biom.t.u-tokyo.ac.jp}
\affiliation{Department of Information and Communication Engineering, Graduate School of Information Science and Technology, The University of Tokyo, Tokyo 113-8656, Japan}
\author{Yoshihiko Hasegawa}
\email{hasegawa@biom.t.u-tokyo.ac.jp}
\affiliation{Department of Information and Communication Engineering, Graduate School of Information Science and Technology, The University of Tokyo, Tokyo 113-8656, Japan}

\begin{abstract}
The Jarzynski estimator is a powerful tool that uses nonequilibrium statistical physics to numerically obtain partition functions of probability distributions. The estimator reconstructs partition functions with trajectories of the simulated Langevin dynamics through the Jarzynski equality. However, the original estimator suffers from slow convergence because it depends on rare trajectories of stochastic dynamics. In this paper, we present a method to significantly accelerate the convergence by introducing deterministic virtual trajectories generated in augmented state space under the Hamiltonian dynamics. We theoretically show that our approach achieves second-order acceleration compared to a naive estimator with the Langevin dynamics and zero variance estimation on harmonic potentials. We also present numerical experiments on three multimodal distributions and a practical example where the proposed method outperforms the conventional method, and provide theoretical explanations.
\end{abstract}

\maketitle

\section{Introduction}
The development of nonequilibrium statistical mechanics has brought about a great many novel algorithms in computational physics and information science \cite{Jarzynski1997,Suwa2010,Ohzeki2010,Rotskoff2019,Habeck2017}. Especially, in recent years, substantial progress has been made in estimation of partition functions \cite{Crooks1999,Habeck2017,Jarzynski1997,Rotskoff2019,Vaikuntanathan2008}. A partition function $Z$ is the normalizing constant of a probability density function $f(x)$ described as a Gibbs distribution:
\begin{align}
    f(x)&=\frac{1}{Z}e^{-\beta E(x)},\\
    \label{eq:partition_function}
    Z&\coloneqq \int e^{-\beta E(x)} \dd x,
\end{align}
where $E(x)$ is the energy of the state $x$ and $\beta$ is inverse temperature. This quantity is fundamental in both natural and information sciences because it characterizes a system in equilibrium: In statistical mechanics, it corresponds to free energy that describes the stability of the system \cite{Reif2009}, and, in machine learning, it is known as the model evidence, which gives an indicator to quantify the likelihood of models for observed data  \cite{MacKay2003,Jaynes2003}. 
Moreover, the partition function in Eq.~\eqref{eq:partition_function} can be regarded as a moment generating function, which gives moments of a system. Therefore, it is important to efficiently calculate the partition function of a system.

However, it is challenging to obtain the partition function or the free energy of a given system. In terms of thermodynamics, the free energy difference between the initial states and final states is determined by the work exerted on the system during a quasi-static process, which is infeasible in finite time \cite{Landau1980}. In terms of numerical computation, calculating the partition function in Eq.~\eqref{eq:partition_function} requires numerical integration over the entire conformation space, which is intractable, especially when the space has many dimensions. 

The remarkable equality demonstrated by Jarzynski \cite{Jarzynski1997} makes it possible to obtain partition functions efficiently. The Jarzynski equality relates the free energy difference $\Delta F$ to work $W$ defined by the trajectories of states in a nonequilibrium process, which is performed by changing the system's configuration in finite time. The equality is given by
\begin{equation}
\expval{e^{-\beta W}}=e^{-\beta\Delta F} = \frac{Z}{Z_0},
\label{eq:JE_def}
\end{equation}
where $Z_0$ is the partition function of the initial distribution and the notation $\langle\cdot\rangle$ is the expectation taken over all possible trajectories during the process. Equation~\eqref{eq:JE_def} holds for arbitrary dynamics as long as the initial states are in equilibrium. From Eq.~\eqref{eq:JE_def}, we can estimate $Z/Z_0$ as an ensemble average of the exerted work. Using this equality, for example, the free energy profile for a molecule was experimentally obtained by repeatedly pulling the molecule with laser optical tweezers \cite{Hummer2010}. Note that since $Z_0$ is rarely accessible in physical systems, we often obtain relative value $Z/Z_0$ to some reference value $Z_0$, rather than $Z$.

An estimator of $Z$ or $Z/Z_0$ can be calculated by simulating the nonequilibrium process \cite{Ritort2008, Neal2001}, and we focus on process modeled by the overdamped Langevin dynamics in this work, because the dynamics is commonly used for systems where the energy function $E(x)$ of the target distribution is determined only by position, which is the case in molecular dynamics simulation or machine learning. Hence, the estimator using the Jarzynski equality for Langevin dynamics is described as the Jarzynski estimator throughout this paper.

In practice, however, convergence of the naive Jarzynski estimator is notoriously slow \cite{Jarzynski2006,Gore2003}.  That is because work $W$ has a large variance and the rare negative value of $W$, which significantly contributes to the estimator, cannot be efficiently sampled. To overcome such a problem, many methods have been proposed. For example, Refs. \cite{Yang2016, Wan2016} devised initial states to alleviate sample bias by allowing almost any distribution for initial states using linear equations. References \cite{Habeck2017, Arrar2019} used a backward process, in which the work distribution can be used to estimate the partition function through Crook's fluctuation theorem \cite{Crooks1999}. Despite such important advances, however, adopting the Langevin dynamics still has a convergence problem: The variance of the estimator remains even if the system is controlled by the optimal protocol unless the distributions of the initial and final states are identical \cite{Schmiedl2007,Zhang_2020}. To overcome such inevitable variance, the authors in Ref.~\cite{Vaikuntanathan2008} introduced an additional flow field and transformed dynamics. They showed that the method achieves zero variance estimation, which means one can obtain the true partition function with only one trajectory, given an ideal flow field, although such a flow field seems impossible to obtain in practical cases.

In this paper, we address the problem of the inevitable error of Jarzynski estimators and significantly improve the convergence in fundamental cases. The key idea of our work is converting the stochastic process described by the Langevin dynamics to a deterministic virtual one. Specifically, we introduce auxiliary momentum and employ the Hamiltonian dynamics to simulate a target system initially connected to a heat bath. In contrast to simulating all particles constituting a system and a heat bath with the Hamiltonian dynamics, the dynamics employed in our method is virtual, viz. not physical, because we transform a system initially coupled to a heat bath to an isolated one and treat particles only in the system.

Our theoretical analyses reveal that our method achieves second-order acceleration with respect to the duration of simulated dynamics, and, furthermore, achieves zero variance estimation at some conditions with harmonic potentials. This property is realized when we employ parallel transport and scaling of harmonic potentials, which are models of a moving laser trap and the time-dependent strength of the trap, respectively \cite{Schmiedl2007}. Moreover, the proposed method mitigates variance when the peaks of the target distribution are far from those of the initial distribution. We conduct numerical experiments on four model systems for which the proposed method outperforms the conventional one and discuss the results theoretically.

\section{Method}
\begin{figure*}[t]
    \begin{minipage}[b]{0.45\textwidth}
        \subfloat[\label{fig:method_HJE}]{
        \includegraphics[width=0.95\linewidth]{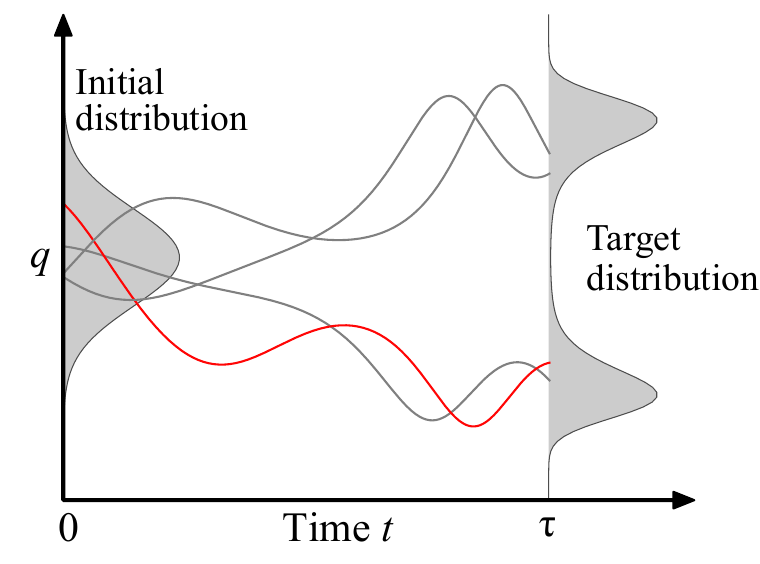}
        }
    \end{minipage}
    \begin{minipage}[b]{0.45\textwidth}
        \subfloat[\label{fig:mehthod_LJE}]{
        \includegraphics[width=0.95\linewidth]{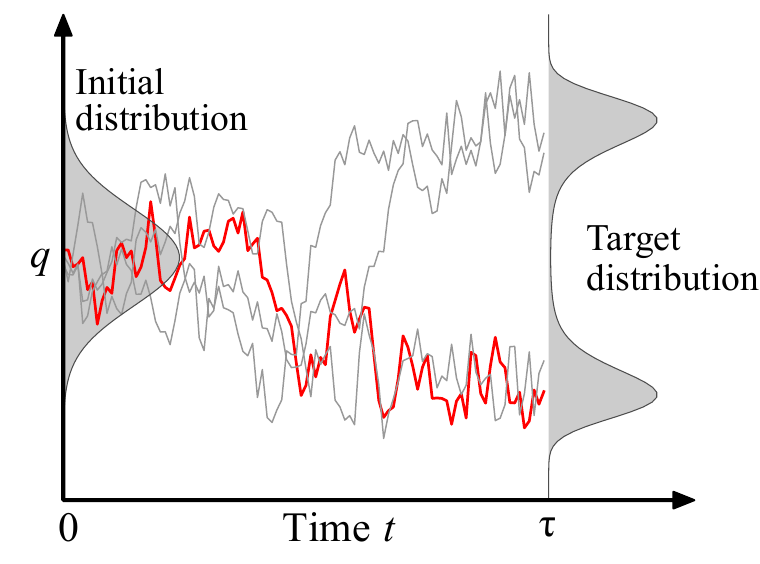}
        }
    \end{minipage}
\caption{Illustrations of (a) the proposed HJE method and (b) the conventional LJE method for estimating the partition function of a target distribution. Both methods generate trajectories, which are depicted by solid curves, whose initial positions are given by some initial distribution. A single trajectory for each method is colored in red for visibility. A protocol for each process is defined during time $t\in[0,\tau]$, and boundary values of the protocol are determined by the initial distribution and the target distribution. The main feature of the HJE is that its trajectories are generated by the Hamiltonian dynamics with virtual momentum, which is deterministic given the initial states, while the LJE adopts the Langevin dynamics, which is stochastic.}
\label{fig:method}
\end{figure*}

Given a potential energy $U_\mathrm{end}(q)$, we aim to calculate the partition function $Z$ defined by
\begin{align}
    Z\coloneqq\int{e^{-\beta U_\mathrm{end}(q)} \dd{q}},
\end{align}
where $q$ is a position. Hereafter, we assume $\beta=1$ without loss of generality. The equilibrium probability distribution corresponding to $U_\mathrm{end}(q)$ is $f_\mathrm{end}^\mathrm{eq}(q)=\exp(-U_\mathrm{end}(q))/Z$. First, we select a probability distribution $f_\mathrm{init}^\mathrm{eq}(q)$ which is arbitrary. Then, to connect these distributions, let us consider a nonequilibrium process during time $t\in[0,\tau]$ characterized with a time-varying potential function $U(q;t)$ with boundary conditions $U(q;0)=U_\mathrm{init}(q)\; \mathrm{and}\; U(q;\tau)=U_\mathrm{end}(q)$. The process is also arbitrary as long as the boundary conditions are satisfied.

In the naive Jarzynski estimator based on the Langevin dynamics \cite{Ritort2008}, which we call the Langevin Jarzynski estimator (LJE) throughout this paper, we consider a trajectory $q(t)$ governed by the overdamped Langevin dynamics:
\begin{align}
    \label{eq:overdamped-Langevin}
    \dv{q}{t}=-\pdv{U(q;t)}{q}+\sqrt{2}\xi(t),
\end{align}
where $\xi(t)$ is Gaussian noise such that its autocorrelation function is a delta function $\expval{\xi(t)\xi(0)}=\delta(t)$. Then, we define the work exerted during a process as
\begin{align}
    \label{eq_work_definitioin}
    W\coloneqq\int_0^{\tau}{\pdv{U(q;t)}{t} \dd{t}}.
\end{align}
The Jarzynski equality in Eq.~\ref{eq:JE_def} holds as an exponential ensemble average of the work of trajectories generated using Eq.~\eqref{eq:overdamped-Langevin}. We obtain an estimator of $Z/Z_0$ by approximating the left-hand side of the Jarzynski equality with a finite number of samples. However, we need a significant number of trajectories to obtain an accurate result because work $W$ has a large variance. 

In the proposed method, which we call the Hamiltonian Jarzynski estimator (HJE), we augment the state space with virtual momentum $p$ which has the same number of dimensions with $q$. Along with $p$, we introduce a kinetic energy with a virtual time-dependent mass $m(t)$:
\begin{align}
    K(p;t)=\dfrac{\norm{p}_2^2}{2m(t)},
\end{align}
where $\norm{\cdot}_2$ is the Euclidean norm. Then, we define the Hamiltonian of the system as
\begin{align}
\label{eq:hamiltonian}
    H(q,p;t)\coloneqq U(q;t)+K(p;t).
\end{align}
Here the schedules of the Hamiltonian $H$ and the mass $m(t)$ are arbitrary as long as the boundary conditions of $U(q;t)$ are satisfied as well as the LJE.

Then, we consider a trajectory $(q(t),p(t))$ under the Hamiltonian dynamics
\begin{align}
    \dv{q}{t}&=\pdv{H(q,p;t)}{p},\\
    \dv{p}{t}&=-\pdv{H(q,p;t)}{q}
\end{align}
and define work consistent with Eq.~\eqref{eq_work_definitioin} \cite{Jarzynski1997}:
\begin{align}
    W&\coloneqq\int_0^{\tau}{\pdv{H(q,p;t)}{t} \dd{t}}\nonumber\\
    &=H(q(\tau),p(\tau);\tau)-H(q(0),p(0);0).
\end{align}
The second equation holds because the entire system is considered isolated throughout the process; that is, there is no heat dissipation. Finally, we can apply the Jarzynski equality for the set of trajectories and obtain the partition function $Z$ with respect to $U_\mathrm{end}(q)$:
\begin{align}
    \label{eq_Jarzynski_Hamiltonian}
    \expval{e^{-W}}
    =\dfrac{\int{e^{-\beta H(q,p;\tau)}}\dd q\dd p}{\int{e^{-\beta H(q,p;0)}}\dd q\dd p}
    =\dfrac{m(\tau)^\frac{N}{2}Z}{m(0)^\frac{N}{2} Z_0},
\end{align}
where $N$ is the dimension of $p$.
The second equation holds because of Eq.~\eqref{eq:hamiltonian}. As well as the LJE, we can estimate $Z$ by approximating the expected value on $W$ in Eq.~\eqref{eq_Jarzynski_Hamiltonian} with an ensemble average of finite samples.

Because the HJE adopts deterministic trajectories described by ordinary differential equations (ODE), we can take advantage of efficient and accurate ODE solvers. This is superior to the case of the LJE, where stochastic differential equations have to be solved. Illustrations of the HJE and the LJE are shown in Figs.~\ref{fig:method_HJE} and \ref{fig:mehthod_LJE}, respectively.

Our approach is inspired by the Hamiltonian Monte Carlo method \cite{Duane1987}. This is an efficient sampling algorithm for accelerating transitions in state space by introducing auxiliary momentum and the Hamiltonian dynamics to replace the Langevin dynamics in the Metropolis-Hastings method \cite{Bishop2006}. In related work, another method introduced auxiliary momentum and the underdamped Langevin dynamics for estimating partition functions \cite{SohlDickstein2011}, but the trajectories in that method are still stochastic and therefore cannot take advantage of deterministic trajectories, in contrast to ours.

We note that the formation of the HJE is different from that based on the Hamiltonian dynamics which traces all particles contained in the system and the connected heat bath. The HJE is also different from the methods that accurately simulate systems connected to thermostats, such as the Nos\'{e}-Hoover method \cite{hoover1985}.

\section{Theoretical analysis}
\label{section:theoreticalAnalysis}
Before testing the proposed method in practical examples, we analytically calculated the properties of the HJE in linear systems, where the partition functions are known and the dynamics is analytically solvable. In particular, we focused on parallel transport, rotation, and scaling of a potential because linear systems are mainly composed of such transformations. 

We compare the LJE and HJE in terms of their variance. To analyze the variance of the estimators, we evaluate the variance of work as a proxy using dissipative work $W_\mathrm{diss}$ \cite{Gore2003}, which is more feasible. $W_\mathrm{diss}$ is defined by
\begin{align}
\label{eq:w_diss}
    W_\mathrm{diss}\coloneqq\expval{W}-\Delta F,
\end{align}
where $\expval{W}$ is average work over all possible trajectories and $\Delta F$ is the difference of free energy between boundary distributions $f_\mathrm{init}^\mathrm{eq}$ and $f_\mathrm{end}^\mathrm{eq}$ satisfying $\Delta F=-\log(Z/Z_0)$. Then we use the following equation as the first-order approximation of the variance of $W$, denoted by $\llrr{W}$ \cite{Gore2003}:
\begin{align}
    \llrr{W} \approx 2W_{\rm diss}.
\end{align}

Moreover, we approximate the distributions during time development with a Gaussian distribution when the initial and final distribution of a protocol are Gaussian. This approximation is valid when the system is near equilibrium during the process.

\subsection{Parallel transport}
\label{section:ParallelTransport}
First, we explore the parallel transport of a one-dimensional (1D) harmonic potential. This is a model of dragged Brownian motion, such as a molecule trapped and pulled by an optical laser \cite{Schmiedl2007,Bustamante2005}. We let the protocol during $t\in[0,\tau]$ be
\begin{gather}
    U(q;t)=\dfrac{\qty(q-\dfrac{\mu_\tau}{\tau}t)^2}{2\sigma^2},
\end{gather}
where $\mu_\tau$ is the transportation distance and $\sigma$ is the standard deviation. This protocol means the harmonic potential is dragged from the origin to $q=\mu_t$ at a constant speed. The corresponding partition functions are $Z=Z_0=\sqrt{2\pi\sigma^2}$. Then, we let the mass during the protocol be constant,
\begin{align}
    m(t)=m_0.
\end{align}
With this setting, we can analytically obtain $W_\mathrm{diss}$ for the LJE and HJE, which are respectively given by
\begin{align}
    \label{eq:paralleltransport_wdiss_LJE}
    W_\mathrm{diss}=\frac{\mu_\tau^2}{\tau}-\dfrac{\mu_\tau^2\sigma^2}{\tau^2}\qty(1-e^{-\dfrac{\tau}{\sigma^2}})
\end{align}
and
\begin{align}
    \label{eq:paralleltransport_wdiss_HJE}
    W_\mathrm{diss}=\dfrac{m_0\mu_\tau^2}{\tau^2}\qty(1-\cos{\dfrac{\tau}{\sqrt{m_0}\sigma}}).
\end{align}
Derivations of Eqs.~\eqref{eq:paralleltransport_wdiss_LJE} and \eqref{eq:paralleltransport_wdiss_HJE} are shown in Appendices \ref{appendix:parallel_Wdiss_LJE} and \ref{appendix:parallel_Wdiss_HJE}, respectively. Note that the result of Eq.~\eqref{eq:paralleltransport_wdiss_LJE} is consistent with a previous analysis of dissipative work in Ref. \cite{Horowitz2009}.

\begin{figure*}[t]
\begin{minipage}[b]{0.45\textwidth}
    \subfloat[\label{fig:paralellTransport}]{
        \includegraphics[width=\linewidth]{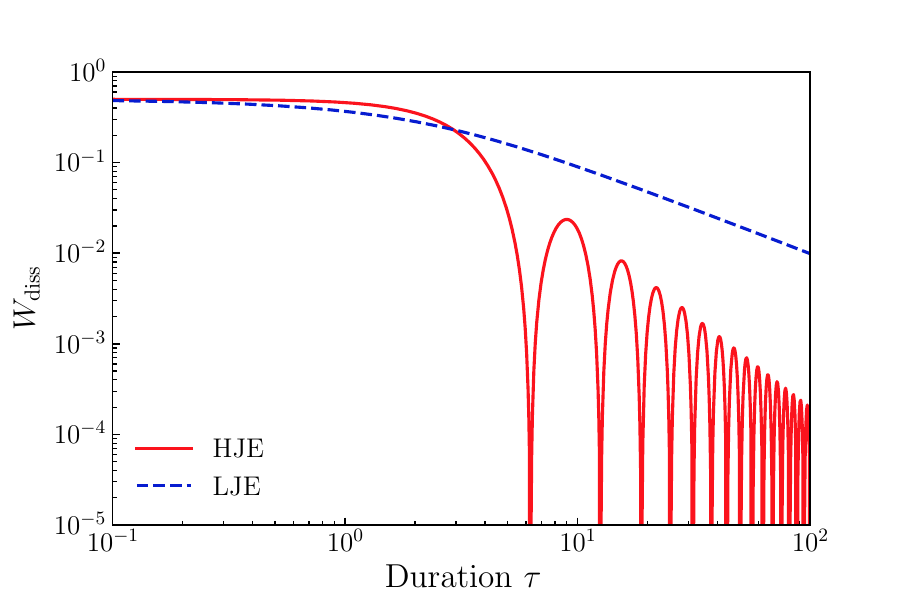}
    }
\end{minipage}
\begin{minipage}[b]{0.45\textwidth}
    \subfloat[\label{fig:var}]{
        \includegraphics[width=\linewidth]{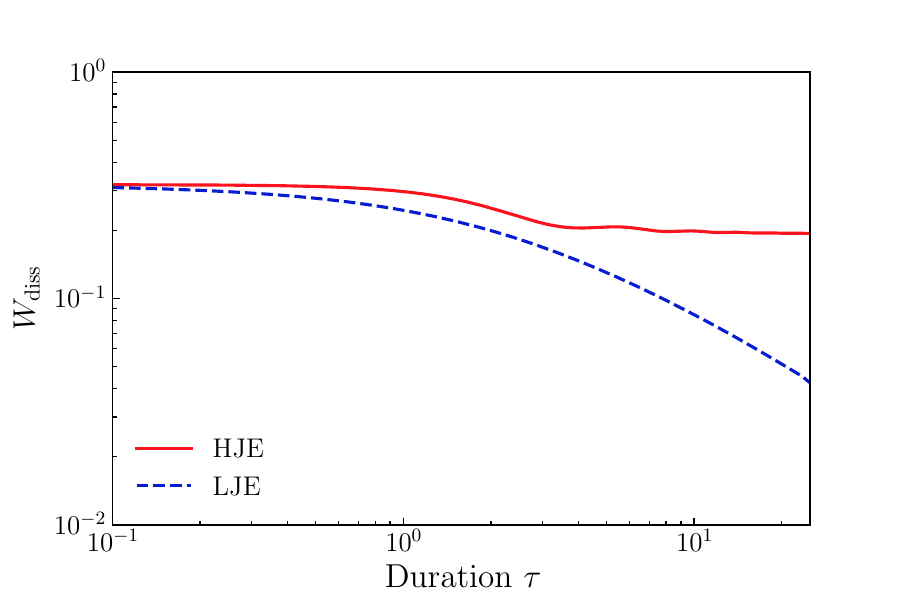}
    }
\end{minipage}
\begin{minipage}[b]{0.45\textwidth}
    \subfloat[\label{fig:var2}]{
        \includegraphics[width=\linewidth]{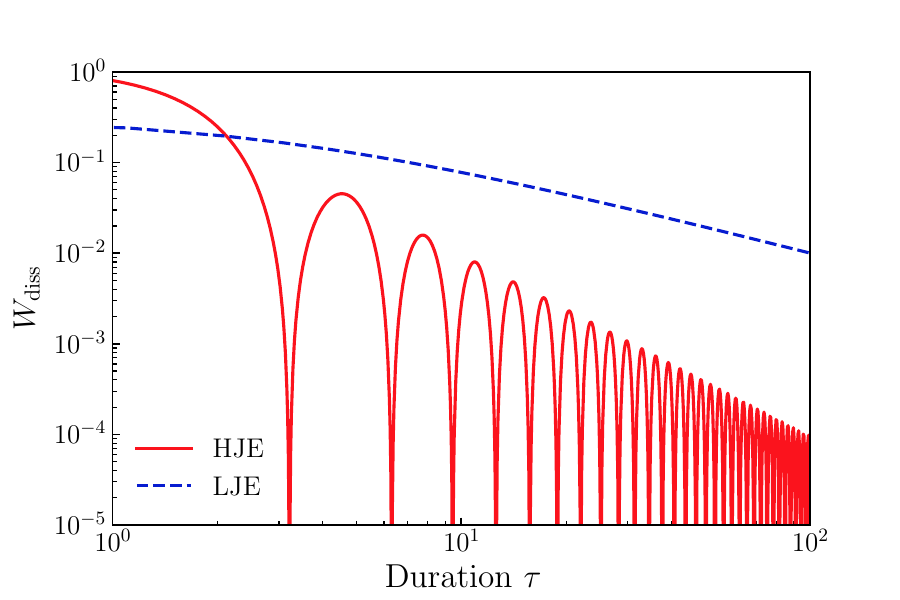}
    }
\end{minipage}
\begin{minipage}[b]{0.45\textwidth}
    \subfloat[\label{fig:rotation}]{
    \includegraphics[width=\linewidth]{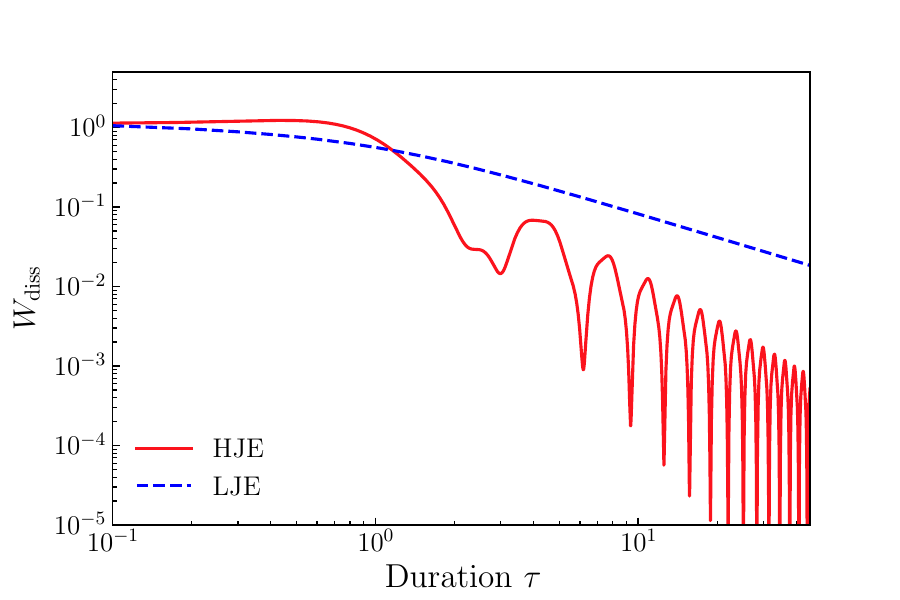}
    }
\end{minipage}
\caption{
Analytically obtained total dissipative work $W_{\rm diss}$ of the HJE and LJE as a function of duration $\tau$ in linear systems. $W_{\rm diss
}$ of the HJE and LJE are depicted by a red solid curve and a blue dashed curve, respectively. (a) Parallel transport of a harmonic potential, where $\sigma^2=1$, $\mu_\tau=1$, and $m_0=1$. (b) Scaling of a harmonic potential, where $\sigma^2_0=1$, $\sigma^2_\tau=4$, and $m(t)=1$. (c) Scaling of a harmonic potential with time-dependent mass for the HJE, where $\sigma^2_0=1$, $\sigma^2_\tau=4$, and $\alpha=1$. (d) Rotation of a harmonic potential, where $a=8$, $b=2$, and $m_0=1$. As a whole, the specific parameters do not change the results qualitatively.
}
\end{figure*}

Equations~\eqref{eq:paralleltransport_wdiss_LJE} and \eqref{eq:paralleltransport_wdiss_HJE} reveal
prominent features of the HJE: Its $W_\mathrm{diss}$ converges in the second order as a function of the reciprocal of $\tau$, while that of the LJE converges in the first order, and, furthermore, the HJE achieves $W_\mathrm{diss}=0$ at $\tau=(2n+1)\pi\sqrt{m_0}\sigma$ for any $n \in \mathbb{N}$ and $\mu_\tau \in \mathbb{R}$, while the LJE never achieves $W_\mathrm{diss}=0$ with finite $\tau$ \cite{Schmiedl2007}. Therefore $W_{\rm diss}$ of the HJE converges faster than that of the LJE when the duration of protocol $\tau$ is longer. 
We note that in the case of $\tau\rightarrow 0$, both Eqs.~\eqref{eq:paralleltransport_wdiss_LJE} and \eqref{eq:paralleltransport_wdiss_HJE} reduce to $W_{\rm diss}=\mu_\tau^2/(2\sigma^2)$, which is equal to the case of importance sampling \cite{Bishop2006}. Importance sampling is a widely used method to estimate an expected value. We plot Eqs.~\eqref{eq:paralleltransport_wdiss_LJE} and \eqref{eq:paralleltransport_wdiss_HJE} in Fig.~\ref{fig:paralellTransport}, which shows that while $W_{\rm diss}$ of the HJE is larger than that of the LJE when the duration $\tau$ is small, the HJE more significantly reduces its $W_{\rm diss}$ than the LJE as $\tau$ becomes large. 

Additionally, we note that the HJE for parallel transport of multi-dimensional harmonic potentials works as well as it does for 1D because in that case, the dynamics can be decomposed into a set of independent 1D parallel transports.

\subsection{Scaling}
\label{subsection:scaling}
Next, we explore the scaling of a harmonic potential by changing the variance of the corresponding distribution from $\sigma^2_0$ to $\sigma^2_\tau$ through a protocol. This is a model of traps in stochastic systems whose strength varies temporally \cite{Schmiedl2007}. The initial distribution and the target distribution are respectively given by
\begin{align}
    f_{\rm init}^{\rm eq}(q)&=\frac{1}{Z_0}e^{-\tfrac{q^2}{2\sigma_0^2}},\\
    f_{\rm end}^{\rm eq}(q)&=\dfrac{1}{Z}e^{-\tfrac{q^2}{2\sigma_\tau^2}},
\end{align}
where $Z_0=\sqrt{2\pi\sigma_0^2}$ and $Z=\sqrt{2\pi\sigma_\tau^2}$. Therefore, the corresponding potential functions are $U_{\rm init}(q)=q^2/(2\sigma_0^2)$ and $U_{\rm end}(q)=q^2/(2\sigma_\tau^2)$. Then we let a linear protocol during $t\in[0,\tau]$ be
\begin{align}
\label{eq:linear_protocol_for_scaling}
    U(q;t)&=\qty(1-\frac{t}{\tau})U_{\rm init}(q)+\frac{t}{\tau}U_{\rm end}(q)
\end{align}
and let the mass be constant, $m(t)=1$.

We plot $W_{\rm diss}$ of the HJE and LJE in Fig.~\ref{fig:var}. The details of the calculation are shown in Appendices \ref{appendix:Wdiss_LJE_scaling} and \ref{appendix:Wdiss_HJE_scaling}, respectively.  Figure~\ref{fig:var} shows that $W_{\rm diss}$ of the HJE converges to some positive value while that of the LJE converges to zero as a function of $\tau$. This result of the HJE is consistent with the well-known fact that an isolated system will generally not be in equilibrium after some protocol, even if the protocol takes infinite time.

However, we can overcome the limitation by recalling the fact that the mass in our method is virtual and arbitrary. To compensate for the difference of variance $\sigma^2_0$ and $\sigma^2_\tau$, we adopt a time-dependent mass. While such a mass protocol is almost infeasible in reality, we can use it because our mass is virtual.

The initial and final distributions are the same as $f_{\rm init}^{\rm eq}(q)$ and $f_{\rm end}^{\rm eq}(q)$, respectively, in the above analysis. Then, we let the protocol during $t\in[0,\tau]$ be
\begin{align}
\label{eq:exp_protocol_for_scaling}
    U(q;t)&=\frac{q^2}{2\sigma(t)^2},\\
    \sigma(t)^2&=\sigma_0^2e^{\frac{\gamma}{\tau}t},
\end{align}
where $\gamma=\log{\sigma_\tau^2/\sigma_0^2}$. For the protocol of a kinetic energy, let the mass be time-dependent:
\begin{align}
    m(t)=\frac{\alpha}{\sigma(t)^2},
\end{align}
where $\alpha$ is a positive constant.
\textbf{}
In this case, we obtain $W_{\rm diss}$ of the HJE as
\begin{align}
    \label{eq:Wdiss_HJE_scaling_time_dependent_mass}
    W_{{\rm diss}}&=-1+\cos^{2}{\omega\tau}+\left(\frac{\alpha\omega^{2}}{2}+\frac{1}{2\alpha\omega^{2}}+\frac{\alpha\gamma^{4}}{32\omega^{2}\tau^{4}}\right.\nonumber\\&\quad\left.+\frac{\alpha\gamma^{2}}{4\tau^{2}}+\frac{\gamma^{2}}{4\omega^{2}\tau^{2}}\right)\sin^{2}{\omega\tau},\\\omega&=\frac{\sqrt{4\tau^{2}/\alpha-\gamma^{2}}}{2\tau},
\end{align}
where $\tau > \sqrt{\alpha}|\gamma|/2$. Details of the derivation are shown in Appendix \ref{appendix:Wdiss_HJE_scaling}. Note that $W_{\rm diss}$ of the LJE needs numerical integration, whose details are shown in Appendix \ref{appendix:Wdiss_LJE_scaling}. We plot $W_{\rm diss}$ of the HJE and LJE in Fig.~\ref{fig:var2}. As well as for parallel transport, the HJE achieves $W_{\rm diss}=0$ with $\omega\tau=\pi n$ for any $n\in\mathbb{N}$, that is, $\tau=\sqrt{\frac{\gamma^2}{4}+\pi^2n^2}$, and converges faster than the LJE thanks to the virtual mass.

\subsection{Parallel transport and scaling}
\label{subsection:parallelVar}
Next, we investigate protocols that include both parallel transport and scaling. In particular, we compare $W_{\rm diss}$ of the HJE and LJE as the distance of parallel transport changes and the magnitude of scaling increases. The initial and final distributions are $\mathcal{N}(0,\sigma^2)$ and $\mathcal{N}(\mu,\sigma^2)$, respectively, where $\mathcal{N}(\mu,\sigma^2)$ denotes a Gaussian distribution with mean $\mu$ and variance $\sigma^2$.
Then we let the protocol during $t\in[0,\tau]$ be
\begin{align}
\label{eq:parallelVar}
    U(q;t)=\frac{1+a\frac{t}{\tau}}{2\sigma^2}\qty(q-\mu\frac{t}{\tau})^2.
\end{align}
This protocol describes that the potential is dragged from the origin to $x=a$ and its strength is varied at a constant ratio over time. Here, we adopt a constant mass of $m(t)=m_0$.

We plot $W_{\rm diss}$ as functions of $a$ and $\mu$ with fixed $\tau=2\pi$ in Fig.~\ref{fig:parallelVar}. The derivation of $W_\mathrm{diss}$ for the protocol is shown in Appendices \ref{appendix:Wdiss_LJE_paralellVar} and \ref{appendix:Wdiss_HJE_paralellVar}. Figure~\ref{fig:parallelVar1} shows that $W_{\rm diss}$ of the LJE is smaller than that of the HJE when $\mu$ is small, which is consistent with the analysis in Section \ref{subsection:scaling} for the case of $\mu=0$ in Eq.~\eqref{eq:parallelVar}. However, $W_{\rm diss}$ of the HJE is significantly smaller than that of the LJE when $\mu$ is large. In addition, as described in Section  \ref{section:ParallelTransport}, $W_{\rm diss}$ is zero for the HJE when $a=0$. This result suggests that as the dragged distance $\mu$ becomes larger, the superiority of the HJE increases, even when the initial distribution and the final distribution cannot be connected only by parallel transport.

\begin{figure}[t]
\centering
\includegraphics[width=0.9\linewidth]{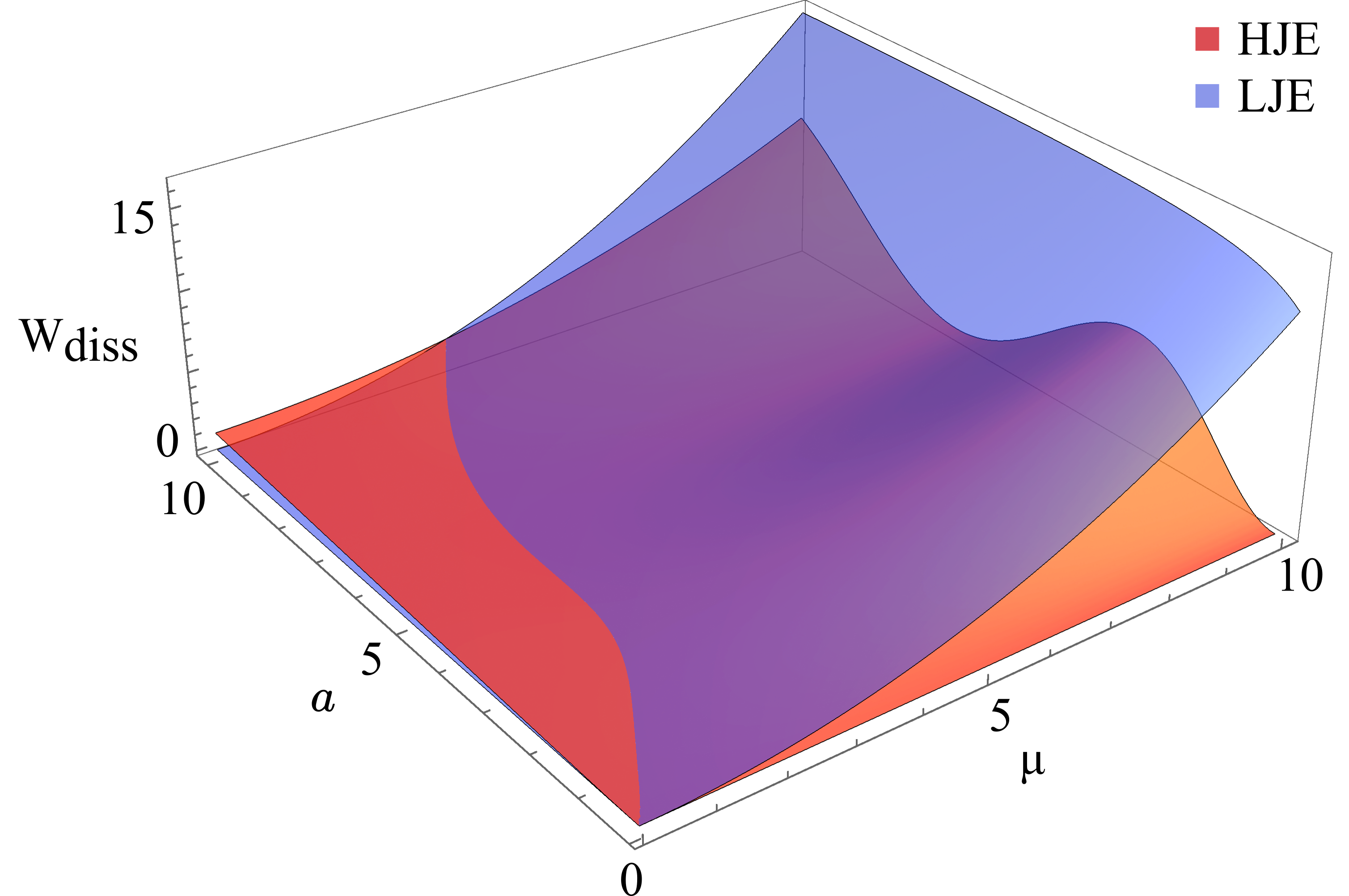}
\caption{\label{fig:parallelVar1}Total dissipative work $W_{\rm diss}$ exerted from a harmonic potential that is dragged and scaled taking duration $\tau=2\pi$, where the HJE and LJE are depicted by red and blue surfaces, respectively. $\mu$ is the dragged distance and $a$ characterizes how much the potential is scaled.
}
\label{fig:parallelVar}
\end{figure}

\subsection{Rotation}
\label{section:Rotation}
Finally, we employ a model of rotation protocol. The initial distribution $f_{\rm init}^{\rm eq}$ and the target distribution $f_{\rm end}^{\rm eq}$ are 
\begin{align}
    f_{\rm init}^{\rm eq}(x,y)&=\dfrac{1}{Z_0}e^{-(ax^2+by^2)},\\
    f_{\rm end}^{\rm eq}(x,y)&=\dfrac{1}{Z}e^{-(bx^2+ay^2)},
\end{align}
where $a$ and $b$ are positive values characterizing the variance of $x$ and $y$, respectively. These distributions are equal to each other through $\pi/2$ rotation. The corresponding partition functions are $Z=Z_0=\pi/\sqrt{ab}$. 

Then, we let a potential function during $t\in[0,\tau]$ be
\begin{align}
    \label{eq:rotation_U)}
    U(x,y;t)=ax'(t)^2+by'(t)^2,
\end{align}
where $(x',y')$ is rotation of $(x,y)$:
\begin{align}
    \mqty(x'(t)\\y'(t))= \mqty(\cos\theta(t) & -\sin\theta(t)\\\sin\theta(t) & \cos\theta(t))\mqty(x\\y)
\end{align}
with $\theta(t)=\pi t/(2\tau)$.

Under this configuration, we can calculate $W_\mathrm{diss}$ analytically, and its derivation is shown in Appendices~\ref{appendix:Wdiss_LJE_rotation} and \ref{appendix:Wdiss_HJE_rotation}. We plot $W_{\rm diss}$ as a function of $\tau$ in Fig.~\ref{fig:rotation}. From Fig.~\ref{fig:rotation}, we find that although $W_{\rm diss}$ of the LJE decreases monotonically, that of the HJE oscillates and its local minima are much smaller than $W_{\rm diss}$ of the LJE at the same $\tau$. Therefore, if we can choose a $\tau$ value where $W_{\rm diss}$ of the HJE is minimal, then it converges significantly faster than the LJE as number of trajectories increases.

\section{Numerical experiments}
In this section, we demonstrate the effectiveness of the HJE through four numerical experiments on model systems where the LJE works poorly. We begin with a toy model to clarify the efficacy of the proposed method. Then, we show the robustness of our method under conditions where the conventional method fails, and verify that the HJE outperforms the LJE for a distribution with a large number of peaks. In addition, we conduct a polymer stretching experiment to clarify the meaning of the virtual trajectories in our method. The result of each experiment is explained based on the theoretical analysis given in the previous section.

\subsection{1D double-well potential}
\label{section:experiment-1D-doublewell-potential}
In this experiment we validate the efficiency of the HJE through a simple but suggestive model. The target distribution is characterized by a double-well potential function,
\begin{align}
    f_{\rm end}^{\rm eq}(q)=\frac{1}{Z}e^{-k q^4+q^2},
\end{align}
where $k$ is a positive parameter. Double-well potentials are widely used to describe bistable states in stochastic systems \cite{Jun2014,Retzker2008}. We use the standard Gaussian distribution as an initial distribution $f_{\rm init}^{\rm eq}$. Then, we construct a protocol for the potential energy during $t\in[0,\tau]$ by linearly interpolating the boundaries:
\begin{align}
    U(q;t)=\qty(1-\dfrac{t}{\tau})\dfrac{q^2}{2}+\dfrac{t}{\tau}\qty(k q^4-q^2).
\end{align}

\begin{figure}[t]
\includegraphics[width=0.45\textwidth]{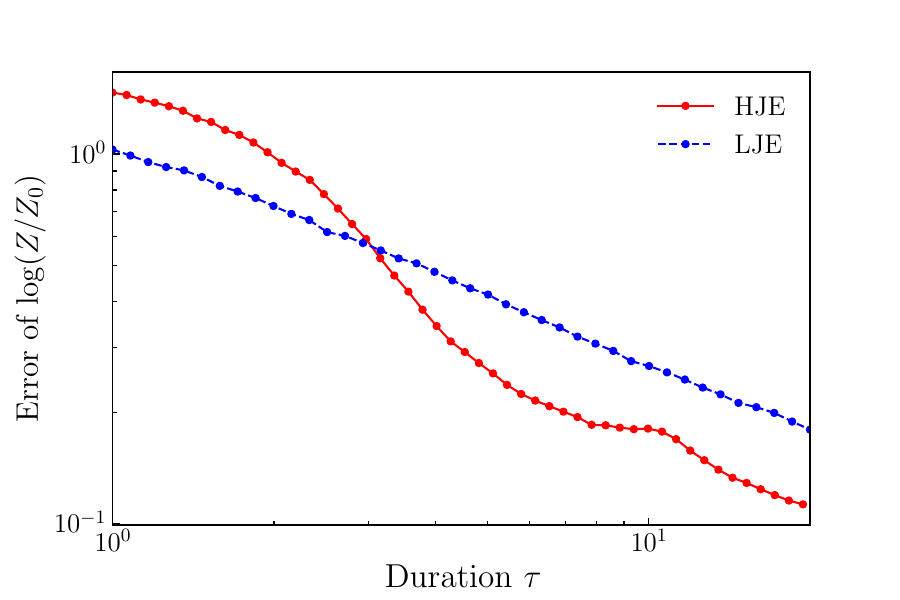}
\caption{\label{fig:twoPeak-result} Error of the estimated partition function for the HJE and LJE, which are depicted by a red solid curve with dots and a blue dashed curve with dots, respectively, as a function of $\tau$ for a 1D double-well potential. The target distribution is characterized by $k=1/16$ and its partition function is $Z=146.372$, which is numerically confirmed. We generated 10 trajectories at each $\tau$ to estimate the partition function and repeated the procedure 100,000 times for both the HJE and LJE. The adopted numeric schemes are the Runge--Kutta 4th-order method for the HJE and the Euler--Maruyama method \cite{kloeden1992} for the LJE. Applying a higher order schemes for stochastic differential equations, order 1.0 strong stochastic Runge--Kutta method \cite{Robler2010}, for the LJE makes no qualitative difference to the result (not shown). We set the time step for numerical integration to be $\Delta t=10^{-3}$.}
\end{figure}

We plot the error of the estimators of $Z/Z_0$ as a function of $\tau$ in Fig.~\ref{fig:twoPeak-result}, which shows that, although the LJE has an error smaller than that of the HJE when $\tau$ is small, the HJE has substantially lower error than that of the LJE as $\tau$ becomes large. Therefore, it is suggested that if we adopt a sufficiently large duration $\tau$, we can estimate the partition function of the target distribution more efficiently by the HJE. Moreover, the error of the HJE oscillates moderately as a function of $\tau$. These results can be explained by the theoretical analysis in the previous section. The protocol of this experiment is approximated with two independent parallel transports of a harmonic potential: one a harmonic potential moving from the origin to the right peak of the target distribution and the other one moving from the origin to the left peak of the target distribution. Then, if the protocol is parallel transport, the error of the HJE oscillates and decreases faster than that of the LJE as a function of $\tau$, which is shown in Section \ref{section:ParallelTransport}.

\subsection{Gaussian mixture model}
\label{section:gaussian_mixture_model}
Next, we examine the robustness of the proposed method against the arrangement of peaks in a target distribution through a two-dimensional (2D) bimodal distribution. In particular, we explore a mixture of two Gaussian distributions with various arrangements. The partition functions for the Gaussian mixture models are $Z=1$ regardless of specific parameters, which is appropriate for comparing the results for target distributions with different parameters.

\begin{figure*}[t]
    \begin{minipage}[b]{0.45\textwidth}
        \subfloat[\label{fig:movingGMM_distribution}]{\includegraphics[width=\linewidth]{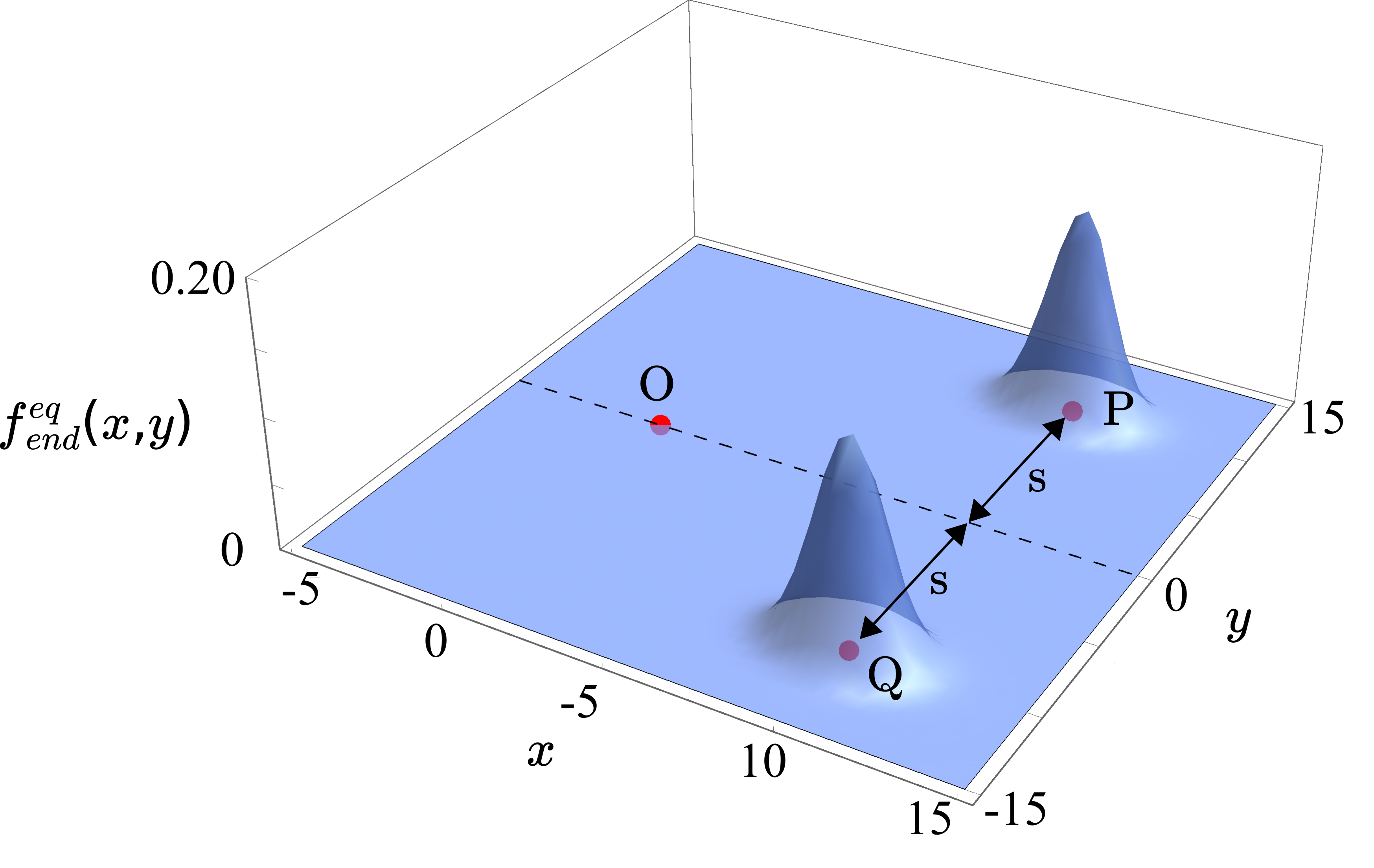}}
    \end{minipage}
    \begin{minipage}[b]{0.45\textwidth}
        \subfloat[\label{fig:movingGMM_result}]{\includegraphics[width=\linewidth]{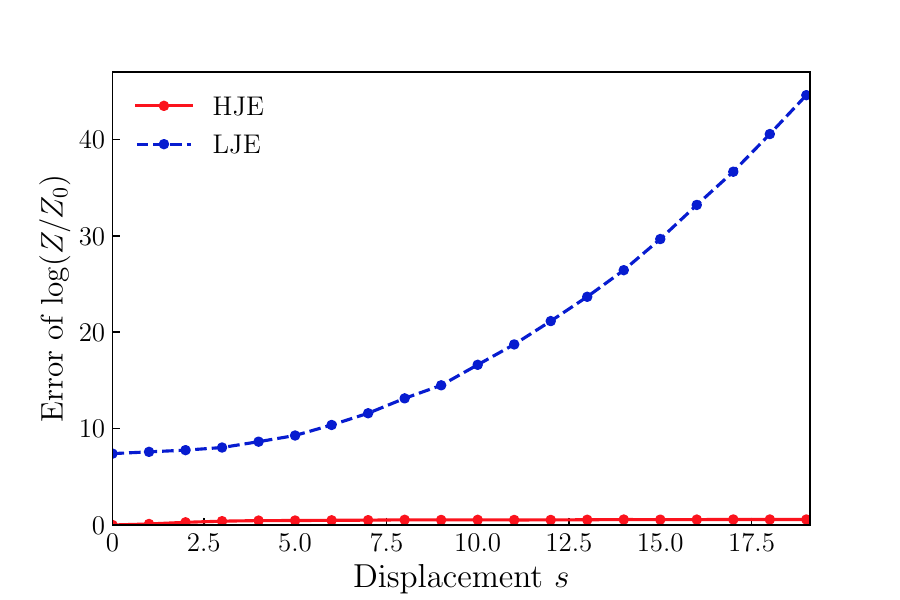}}
    \end{minipage}
\caption{Illustration of the target distribution and error of the estimated partition function. (a) Plot of the target distribution. The distribution is a mixture of Gaussian distributions, and the peaks are located at points ${\rm P}=(a,s)$ and ${\rm Q}=(a,-s)$. We define the displacement $s$ given by the distance between the x-axis and the peaks of the target distribution so as to characterize their arrangement. (b) Error for the HJE and LJE, which are depicted by a red solid curve with dots and a blue dashed curve with dots, respectively, as a function of $s$, where $\tau=2\pi$, $\sigma^2=1$, and $a=10$. We generated 10 trajectories at each $s$ to estimate the partition function and repeated the procedure 1,000 times for the HJE and LJE. The settings for the numerical schemes are the same as in Section \ref{section:experiment-1D-doublewell-potential}.
}
\end{figure*}

The target distribution is a mixture of Gaussian distributions $\mathcal{N}(\mu_P,\sigma^2)$ and $\mathcal{N}(\mu_Q,\sigma^2)$ in state space $q=(x,y)$, where $\mu_P=(a,s)$ and $\mu_Q=(a,-s)$. That is,
\begin{align}
\label{eq:2d_GMM}
f_{\rm end}^{\rm eq}(x,y)&=\frac{1}{Z}\exp\qty(-U_{\rm end}(x,y)),\\
\nonumber
U_{\rm end}(x,y)&=-\log\left(\frac{1}{4\pi\sigma^2}e^{-\frac{(x-a)^2+(y-s)^2}{2\sigma^2}}\right.\\
&\quad+\left.\frac{1}{4\pi\sigma^2}e^{-\frac{(x-a)^2+(y+s)^2}{2\sigma^2}}\right).
\end{align}
As shown in Fig.~\ref{fig:movingGMM_distribution}, the distribution given by Eq.~\eqref{eq:2d_GMM} is bimodal, and its peaks are located at points $P=\mu_P$ and $Q=\mu_Q$. Hereafter, we define the displacement $s$, which describes how the two peaks are separated from each other. Then we adopt a Gaussian distribution at the origin ${\rm O}$ as the initial distribution, that is, $U_{\rm init}(x,y)=\qty(x^2+y^2)/\qty(2\sigma^2)$, and let the protocol between $t\in[0,\tau]$ be
\begin{align}
    U(x,y;t)&=\qty(1-\frac{t}{\tau})U_{\rm init}(x,y)+\frac{t}{\tau}U_{\rm end}(x,y).
\end{align}

We sweep the displacement $s$ to reveal how the arrangement of the peaks of the distributions affects the HJE and LJE. Moreover, we let $\tau=2\pi\sqrt{\sigma^2}$, where $W_{\rm diss}$ of the HJE in the parallel transport protocol given by Eq.~\eqref{eq:paralleltransport_wdiss_HJE} vanishes.

We plot the error of the estimated $Z/Z_0$ as a function of $s$ in Fig.~\ref{fig:movingGMM_result}, which shows that the HJE has an error significantly smaller than that of the LJE for any displacement $s$. Moreover, although the error of the LJE increases rapidly as a function of $s$, that of the HJE increases only slightly. These results can be explained by the analysis of parallel transport in Section \ref{section:ParallelTransport}. When $s=0$, the target distribution is a Gaussian distribution, as explained by the analysis in Section \ref{section:ParallelTransport}, which demonstrates that the error of the HJE vanishes regardless of the distance between the peaks of the initial and final distributions. This suggests that the significantly small error of the HJE in this experiment is due to the protocol being approximately decomposable into two independent parallel transports: on a harmonic potential moving from ${\rm O}$ to ${\rm P}$ and the other one moving from ${\rm O}$ to ${\rm Q}$. As displacement $s$ grows, the peaks of the target distribution separate from each other and the distance becomes large between the points ${\rm P}$ or ${\rm Q}$, which are means of the peaks, and the origin ${\rm O}$, which is the mean of the initial distribution. If the protocol can be considered as a parallel transport, the distance between ${\rm O}$ and ${\rm P}$ or ${\rm Q}$ has no effect on the performance of the HJE, while the error of the LJE increases as a function of the distance. These are the reasons why the qualitative difference appears in the experiment shown in Fig.~\ref{fig:movingGMM_result}.

\subsection{Multimodal distribution}
\label{section:multimodal}
In this experiment, we employ a distribution with many peaks to demonstrate the effectiveness of the HJE for multimodal distributions. We let the dimension of the state space $q$ be 16 and the target distribution defined over $q$ is
\begin{align}
\label{eq:coshManyPeaks_distribution}
    f^{\rm eq}_{\rm end}(q)&=\frac{1}{Z}\exp(-U(q;\tau)),\\
    U_{\rm end}(q)&=-\log\left\{\sum_{i=1}^N{\exp\left(-\cosh\|q-\mu_i\|_2\right)}\right\},
\end{align}
where $N$ is the number of the peaks and $\mu_i$ is the location of the $i$th peak.

We use a zero-mean Gaussian distribution with a variance $\sigma^2$ as the initial distribution:
\begin{align}
    \label{eq:coshManyPeaks_f}
    f^{\rm eq}_{\rm init}(q)=\frac{1}{2\pi\sigma^2}\exp\qty(-\frac{\|q\|^2_2}{2\sigma^2}).
\end{align}
Then we let the protocol during $t\in[0,\tau]$ be
\begin{align}
    \nonumber
    U(q;t)&=\qty(1-\frac{t}{\tau})\frac{\|q\|_2^2}{2}\\
    &\quad+\frac{t}{\tau}\left(-\log\left\{\sum_{i=1}^N{\exp\left(-\cosh\left\|q-\mu_i\frac{t}{\tau}\right\|_2\right)}\right\}\right).
\end{align}

\begin{figure}[t]
\includegraphics[width=1\linewidth]{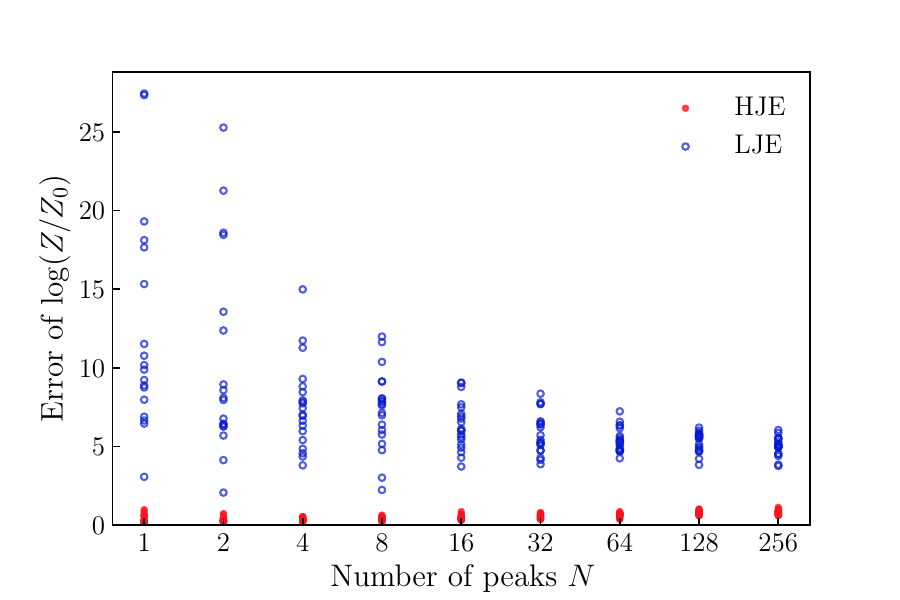}
\caption{\label{fig:coshManyPeaks-result} Error of the estimated partition function for the HJE and LJE, which are depicted by red dots and blue circles, respectively, as a function of $N$ characterizing multimodal distributions in 16-dimensional space. Points at each $N$ corresponds to different sets of $\{\mu_i\}$. The ground truth of $Z$ is numerically obtained as in Appendix \ref{appendix:Z_of_multimodal}. We tested 20 sets of $\{\mu_i\}$. Then, we generated 1,000 trajectories for each set to estimate the partition function, and repeated the procedure 10 times for the HJE and LJE. The duration of process is $\tau=4\pi$. The settings for the numerical schemes are the same as in Section \ref{section:experiment-1D-doublewell-potential}.
}
\end{figure}

We sweep the number of peaks $N$ from 1 to 256 in order to reveal how the performance of the HJE scales as the target distribution has more peaks. For each $N$, we test 20 sets of $\{\mu_i\}_{i=1}^{N}$, 
which are randomly sampled from a Gaussian distribution $\mathcal{N}(0,25)$. 

We plot the error of the estimators of $Z/Z_0$ in Fig.~\ref{fig:coshManyPeaks-result} as a function of $N$, which shows that the error of the HJE is substantially smaller that that of the LJE for all $N$. Roughly speaking, the HJE accurately estimates the order of the ground truth, while the LJE differs by two digits. This result can be explained by the theoretical analysis in Section \ref{section:theoreticalAnalysis} and the previous experiment on a Gaussian mixture model in Section \ref{section:gaussian_mixture_model}. The protocol in this experiment can be approximated by independent parallel transport from the initial distribution to peaks in the target distribution, where the HJE has been demonstrated to perform well. On the other hand, the peak in the initial distribution and those in the target distribution have different shapes in terms of the second or higher order moments, which results in the increment of the variance of the estimator as shown in the theoretical analysis of the scaling protocol in Section \ref{subsection:scaling}. Then we conclude that the reason for the success of the HJE in this experiment is because the contribution of parallel transport is more substantial than deterioration by scaling, which is suggested by the theoretical analysis of the protocol that includes both parallel transport and scaling in Section \ref{subsection:parallelVar}.

\subsection{Stretching polymer chain in a solvent}
\begin{figure}[t]
    \begin{minipage}[b]{0.4\textwidth}
        \subfloat[\label{fig:rouse_model}]{\includegraphics[width=\linewidth]{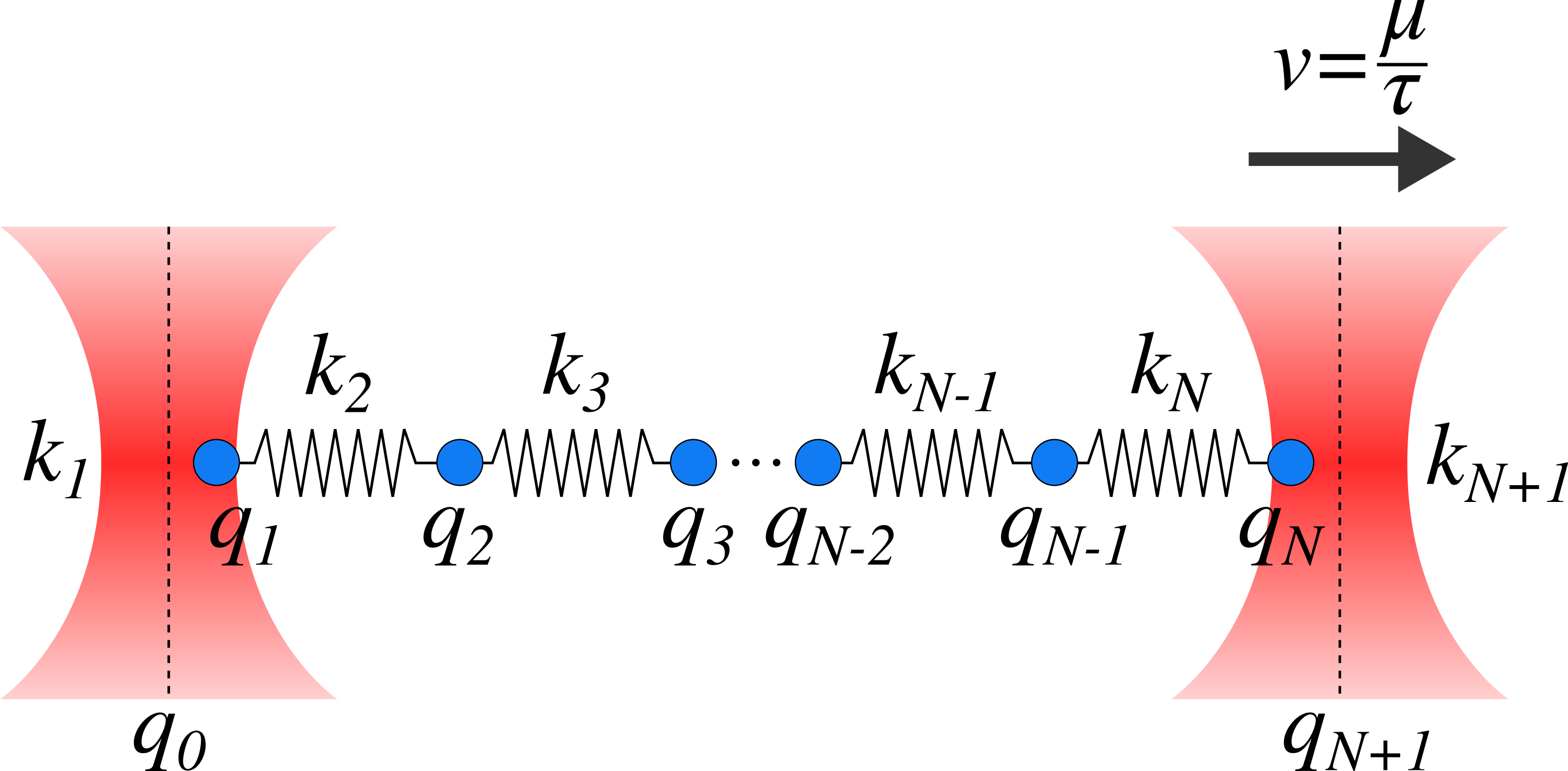}}
    \end{minipage}
    \begin{minipage}[b]{0.45\textwidth}
        \subfloat[\label{fig:rouse_result}]{\includegraphics[width=1\linewidth]{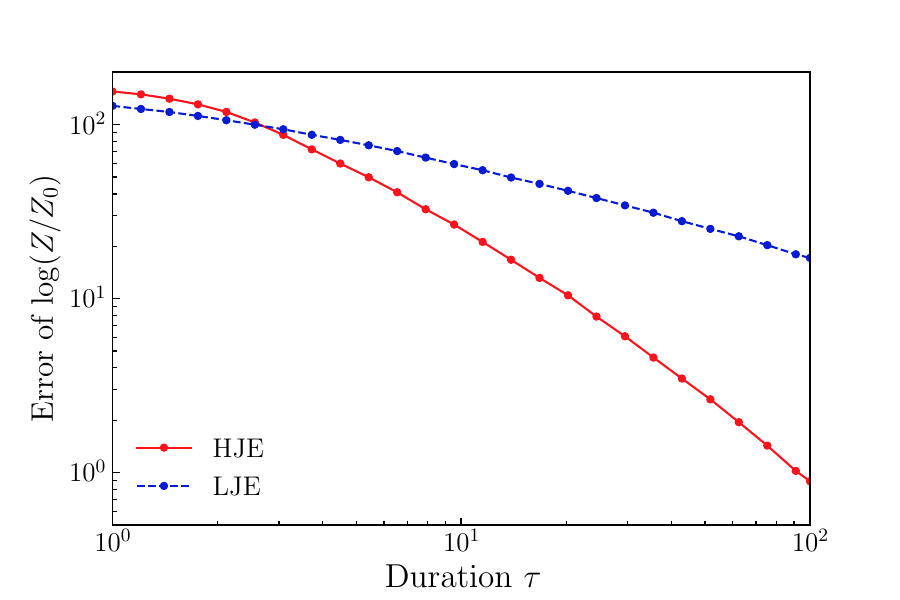}}
    \end{minipage}
\caption{Illustration of stretched Rouse model and error of the estimated partition function. (a) Model description. A rouse model consists of $N$ beads of positions $q_1$ to $q_N$ and harmonic springs connecting them with spring constants $k_2$ to $k_N$. Each end point of the chain is held by optical laser traps with positions $q_0$ and $q_{N+1}$ and spring constants $k_0$ and $k_{N+1}$, respectively. The right trap moves at a constant velocity $v=\mu/\tau$. (b) Error of the estimated partition function with the protocol of the stretching Rouse model for the HJE and LJE, depicted by a red solid curve with dots and a blue dashed curve with dots, respectively, as a function of $\tau$. The model is characterized by $N=100$, $\mu=20$, and $k_n=1$ for $n=1,...,N$. We generated 10 trajectories at each $\tau$ to estimate the partition function and repeated the procedure 2,000 times for both the HJE and LJE. The settings for the numerical schemes are the same as in Section \ref{section:experiment-1D-doublewell-potential}.}
\end{figure}
Finally, we clarify the meaning of the virtual dynamics of our method and its efficacy through an experiment using the stretching Rouse model, which is an ideal polymer chain in a solvent  \cite{rouse1953,strobl1997}. In the model, a polymer chain is split into subsections which show rubber-like behavior. The subsections are considered beads with no volume and connected to adjacent elements with harmonic springs. The beads follow the Langevin dynamics. Note that each bead in the Rouse model corresponds not to a molecule but to a segment of a molecular chain.

We conduct a numerical experiment stretching the chain with two optical traps. One trap is grabs one end point of the chain and is fixed at its initial position, and the other grabs the other end point of the chain and moves to the target position. The interaction between the traps and the grabbed beads are also modeled as harmonic springs \cite{Dhar2005}. To formulate the system, we consider a series of elements: an optical trap, $N$ beads, and the other trap, each connected with springs. We represent the position of the $i$-th element by the scalar value $q_i$: $q_0$ for the first trap, $q_1,...,q_n$ for beads, and $q_{N+1}$ for the other trap. The system is illustrated in Fig.~\ref{fig:rouse_model}.

Note that in the proposed method, the beads are considered to be free particles, which have no physical counterparts because the subsection-dividing coarse-graining in the Rouse model is valid only when the polymer is in a solvent or other similar circumstances, but not in a vacuum.

The potential energy of the chain is given by
\begin{align}
    U(q;t)=\sum_{n=1}^{N+1}\frac{k_n}{2}(q_n-q_{n-1})^2,
\end{align}
where $k_n$ is the strength of the $n$-th spring connecting elements $n$ and $n+1$. 

We consider a linear protocol moving the trap position from $0$ to $\mu$ at a constant velocity $\mu/\tau$. That is, the initial and final distributions are \begin{align}
    f_{\rm init}^{\rm eq}(q) &= \frac{1}{Z_0}\exp(-U(q;0)),\\
    f_{\rm end}^{\rm eq}(q) &= \frac{1}{Z}\exp(-U(q;\tau)),
\end{align}
and the position of the moving trap is described by $q_{N+1}(t)=t\mu/\tau$. The exact solution of $Z$ for this model is characterized by the position of the moving trap $q_{N+1}$:
\begin{align}
    Z(q_{N+1})&=\sqrt{\frac{(2\pi)^n}{\det A}}\exp\qty(-\frac{1}{2}\bar{k}q_{N+1}^2),
\end{align}
where $A$ is a trigonal matrix with elements $A_{n,n}=k_n+k_{n+1}$, $A_{n,n+1}=-k_{n+1}$, $A_{n,n-1}=-k_n$ and $\bar{k}$ is the harmonic mean of $k_n$ satisfying  $1/\bar{k}=\sum_{n=1}^{N+1}1/k_n$ \cite{Dhar2005}. The initial distribution $f_{\rm init}^{\rm eq}(q)$ is a Gaussian distribution with a mean of zero and a covariance matrix $A^{-1}$.

We sweep the duration of the process $\tau$ and plot the error of the estimated $Z/Z_0$ for $N=100$ in Fig.~\ref{fig:rouse_result}, in which the error for the HJE decreases faster and becomes smaller than that for the LJE as $\tau$ increases. Therefore, for large enough $\tau$, the HJE outperforms the LJE in terms of fast convergence.

We now provide a theoretical explanation for the result. The distribution of positions during the time development is Gaussian and its mean and variance are given as follows \cite{Dhar2005}:
\begin{align}
    \expval{q}&=A^{-1}h\\
    \expval{(q-\expval{q})(q^T-\expval{q^T})}&=A^{-1}
\end{align}
where $h^T=(0,0,...,0,k_{N+1}q_{N+1}(t))$ is time dependent. Therefore, the protocol can be considered to represent parallel transport of the peak in a distribution of positional states. Then, as we showed theoretically in Section \ref{section:ParallelTransport}, the HJE achieves faster convergence than the LJE for parallel transport. Thus, the HJE works well for this experiment.

\section{Concluding remarks}
In this paper, we have addressed the problem of slow convergence of Jarzynski estimators for partition functions by introducing deterministic virtual trajectories instead of the original stochastic trajectories. We proved that our method achieves second-order acceleration with respect to the duration of simulated dynamics and furthermore zero-variance estimation in the case of harmonic potentials. Then, we numerically showed that the HJE outperforms the LJE in four examples and provided theoretical explanations why the HJE exhibits better performance in the examples. In short, it is suggested that if the protocol can be approximated by some independent parallel transport protocols, the error of the HJE decreases faster than that of the LJE as the duration $\tau$ grows. We suggest that probability distributions modeling multistable systems would satisfy this requirement.

The HJE uses the Hamiltonian dynamics to make the simulated trajectories deterministic. Theoretical analysis of parallel transport, scaling with a constant mass, and rotation of potentials implies that if the eigenfrequency of the initial distribution is preserved through the process, as in the cases of parallel transport and rotation, each trajectory of the HJE is less dissipative than when the frequency varies, which is the case of scaling. Furthermore, the HJE can compensate for the variation of the eigenfrequency during a process if an appropriate time-dependent mass is employed, as is the case of scaling with a time-dependent mass.

In addition to the above advantages, our work enables the employment of efficient and accurate ODE solvers when simulating the nonequilibrium process. Considering that the LJE needs to solve stochastic differential equations, this property makes the present method easier to apply.

Despite the fact that there are cases where $W_{\rm diss}$ of the HJE does not converge to zero even when the case of $\tau\rightarrow\infty$ like the scaling protocol shown in Section \ref{subsection:scaling}, our work has the potential to be a dramatically efficient estimator of partition functions, especially for multistable systems. The HJE is a fundamental improvement of the LJE, so various future work based on the HJE could be expected to address its limitations, as many efficient methods have been developed based on the LJE. We presume that time-dependent mass, which we used in the scaling protocol in Section \ref{subsection:scaling}, is particularly promising for the HJE. 

\begin{acknowledgments}
This work was supported by the Ministry of Education, Culture, Sports, Science and Technology (MEXT) KAKENHI Grant No.~JP19K12153.
\end{acknowledgments}

\appendix
\section{$W_{\rm diss}$ of the LJE for parallel transport}
\label{appendix:parallel_Wdiss_LJE}
We derive $W_{\rm diss}$ of the LJE for the parallel transport protocol. When a state $q$ of a system satisfies the Langevin dynamics, the time development of the probability distribution of the system $f(q,t)$ is described by following the Fokker--Planck equation \cite{Risken1989} under the condition of inverse temperature $\beta=1$:
\begin{align}
\label{eq:fokker}
    \pdv{f(q,t)}{t}=\pdv{}{q}\left[\pdv{U(q;t)}{q}+\pdv{}{q} \right]f(q,t).
\end{align}
Using Eq.~\eqref{eq:fokker}, we obtain the time derivative of the average and variance of position, which are denoted as $\expval{q(t)}$ and $\llrr{q}$ at time $t$:
\begin{align}
    \label{eq:appendix_langevin_mu}
    \dv{\expval{q(t)}}{t}&=-\int{\pdv{U(q;t)}{q}f(q,t)\dd q},\\
    \label{eq:appendix_langevin_var}
    \dv{\llrr{q(t)}}{t}&=-2\int{q\pdv{U(q;t)}{q}f(q,t)\dd q}+2-2\mu(t)\dv{\mu(t)}{t}.
\end{align}
With the setting in Section \ref{section:ParallelTransport}, by solving Eqs.~\eqref{eq:appendix_langevin_mu} and \eqref{eq:appendix_langevin_var}, we obtain
\begin{align}
    \expval{q(t)}&=\frac{\mu_\tau}{\tau}\qty(t-\sigma^2\qty(1-e^{-\frac{t}{\sigma_0^2}})),\\
    \llrr{q(t)}&=\sigma^2_0.
\end{align}

We can verify that the density state at time $t$ is a Gaussian distribution with mean $\mu(t)$ and variance $\sigma^2(t)$ by substituting $f(q,t)$ into Eq.~\eqref{eq:fokker}.
Finally, with the definitions of work $W$ [Eq.~\eqref{eq_work_definitioin}] and $W_{\rm diss}$ [Eq.~\eqref{eq:w_diss}] we obtain Eq.~\eqref{eq:paralleltransport_wdiss_LJE}.

\section{$W_{\rm diss}$ of the HJE for parallel transport}
\label{appendix:parallel_Wdiss_HJE}
We derive $W_{\rm diss}$ of the HJE for the parallel transport protocol. To obtain the dissipative work $W_{\rm diss}$, we calculate the Kullback--Leibler (KL) divergence between the equilibrium distribution $f_{\rm end}^{\rm eq}$ corresponding to the final Hamiltonian $H(q,p;\tau)$ and the realized distribution $f_{\rm end}$ at the end of the process because the following equality holds \cite{Vaikuntanathan2009}:
\begin{align}
    \label{eq:Wdiss_Dkl}
    W_{\rm diss}
    &=D_{\rm KL}(f_{\rm end}||f_{\rm end}^{\rm eq})\\
    \nonumber
    &\coloneqq\int{f_{\rm end}(q(\tau),p(\tau))\log{\frac{f_{\rm end}(q(\tau),p(\tau))}{f_{\rm end}^{\rm eq}(q(\tau),p(\tau))}}\dd q(\tau)\dd p(\tau)}.
\end{align}
With the setting in Section \ref{section:ParallelTransport}, by solving the Hamiltonian dynamics with an initial condition $(q_0,p_0)$ we have the final state $(q(\tau),p(\tau))$ as
\begin{align}
    q(\tau)&=\qty(\frac{p_0}{\sqrt{m_0}}-\frac{\mu_\tau}{\tau})\sigma\sin\frac{\tau}{\sqrt{m_0}\sigma}+q_0\cos\frac{\tau}{\sqrt{m_0}\sigma}+\mu_\tau,\\
    \nonumber
    p(\tau)&=\qty(\frac{p_0}{\sqrt{m_0}}-\frac{\mu_\tau}{\tau})\cos\frac{\tau}{\sqrt{m_0}\sigma}\\&\quad-\frac{\sqrt{m_0}q_0}{\sigma}\sin\frac{\tau}{\sqrt{m_0}\sigma}+\frac{m_0\mu_\tau}{\tau}.
\end{align}

Then, Liouville's theorem \cite{tuckerman2010}, which describes that the Hamiltonian dynamics preserves the density of states along each trajectory, is used to obtain the final distribution as
\begin{align}
    f_{\rm end}(q(\tau),p(\tau))=f_{\rm init}^{\rm eq}(q_0,p_0).
\end{align}

Finally, we can calculate the KL divergence as
\begin{align}
    \label{eq:Dkl_HJE}
    \nonumber
    &D_{\rm KL}(f_{\rm end}||f_{\rm end}^{\rm eq})\\
    \nonumber
    &=\int{f_{\rm init}^{\rm eq}(q(\tau),p(\tau))\log{\frac{f_{\rm init}^{\rm eq}(q(\tau),p(\tau))}{f_{\rm end}^{\rm eq}(q(\tau),p(\tau))}}\dd q(\tau)\dd p(\tau)}\\
    \nonumber
    &=\int{f_{\rm init}^{\rm eq}(q(\tau),p(\tau))\log{\frac{f_{\rm init}^{\rm eq}(q(\tau),p(\tau))}{f_{\rm end}^{\rm eq}(q(\tau),p(\tau))}}\dd q_0\dd p_0}\\
    &=\frac{m_0\mu_\tau^2}{\tau^2}\qty(1-\cos\frac{\tau}{\sqrt{m_0}\sigma}),
\textbf{}\end{align}
where the second equality holds because the Jacobian for the variable transformation $q(\tau)\rightarrow q_0$ and $p(\tau)\rightarrow p_0$ is 1. With Eqs.~\eqref{eq:Wdiss_Dkl} and \eqref{eq:Dkl_HJE} we obtain Eq.~\eqref{eq:paralleltransport_wdiss_HJE}.

\section{$W_{\rm diss}$ of the LJE for scaling}
\label{appendix:Wdiss_LJE_scaling}
We derive $W_{\rm diss}$ for the scaling protocol. We obtain a differential equation on the variance of position $\llrr{q(t)}$ by substituting $U(q;t)$ given by Eq.~\eqref{eq:linear_protocol_for_scaling} for Eq.~\eqref{eq:appendix_langevin_var}:
\begin{align}
\label{eq:dVardT_for_linearScaling}
    \dv{\llrr{q(t)}}{t}&=\frac{\sigma_\tau^2(\tau-t)+\sigma^2_0t}{\sigma_0^2\sigma_\tau^2\tau}\llrr{q(t)}+2.
\end{align}
Then, the average work exerted during the process is
\begin{align}
\nonumber
\expval{W}&=\expval{\int_0^\tau\pdv{U(q;t)}{t}\dd t}\\
\label{eq:dWdT_for_linear_scaling}
    &=\int_0^\tau{\frac{\sigma_0^2-\sigma_\tau^2}{2\sigma_0\sigma_\tau^2\tau}\llrr{q(t)}}\dd t.
\end{align}
We can calculate Eq.~\eqref{eq:dWdT_for_linear_scaling} by solving Eq.~\eqref{eq:dVardT_for_linearScaling} numerically. Finally, we obtain $W_{\rm diss}$ by subtracting $\Delta F=-\log[\sigma_\tau/\sigma_0]$.

Next, we derive $W_{\rm diss}$ for the scaling protocol given by Eq.~\eqref{eq:exp_protocol_for_scaling}. The derivative of the variance of position is
\begin{align}
\label{eq:dVardT_for_expScaling}
    \dv{\llrr{q(t)}}{t} &= 2(1-e^{-\frac{\gamma t}{\tau}})\llrr{q(t)}.
\end{align}
Then, the average work exerted during the process is
\begin{align}
\label{eq:dWdT_for_exp_scaling}
\expval{W}&=\int_0^\tau{-\frac{\gamma}{2\sigma_0^2\tau}e^{-\frac{\gamma t}{\tau}}\llrr{q(t)}}\dd t.
\end{align}
We can calculate Eq.~\eqref{eq:dWdT_for_exp_scaling} by solving Eq.~\eqref{eq:dVardT_for_expScaling} numerically. Finally, we obtain $W_{\rm diss}$ by subtracting $\Delta F=-\log[\sigma_\tau^2/\sigma_0^2]$.

\section{$W_{\rm diss}$ of the HJE for scaling}
\label{appendix:Wdiss_HJE_scaling}
We derive $W_{\rm diss}$ of the HJE for the scaling protocol. The distribution during the process is not Gaussian in this case, but we approximate the distribution with a Gaussian distribution. This approximation is valid when the system is near equilibrium during the process. 

 When a state $(q,p)$ of a system satisfies the Hamiltonian dynamics the time development of the probability distribution of the system $f(q,p,t)$ is described by Liouville's equation \cite{tuckerman2010}:
\begin{align}
\label{appendix:eq_liouville_equation}
    \pdv{f}{t}+\sum_{i=1}^n\qty(\pdv{f}{q_i}\dv{q_i}{t}+\pdv{f}{p_i}\dv{p_i}{t})=0,
\end{align}
where $n$ is the dimension of $q$ and $p$.

Similar to the case of the Fokker--Plank equation in Appendix \ref{appendix:parallel_Wdiss_LJE}, we use Eq.~\eqref{appendix:eq_liouville_equation} to obtain the time derivatives of states' variance $\llrr{\,\cdot\,}$ and covariance $\llrr{\cdot,\cdot}$ at time t under the protocol given by Eq.~\eqref{eq:linear_protocol_for_scaling}:
\begin{align}
    \label{eq:appendix_hamiltonian_linearScaling_begin}
    \dv{\llrr{q(t)}}{t}&=2\llrr{q(t),p(t)},\\
    \dv{\llrr{p(t)}}{t}&=-2\llrr{q(t), p(t)}\frac{\sigma_\tau^2\tau+(\sigma_0^2-\sigma_\tau^2)t}{\sigma_0^2\sigma_\tau^2\tau},\\
    \nonumber
    \dv{\llrr{q(t), p(t)}}{t}&=\frac{1}{\sigma_0^2\sigma_\tau^2\tau}\left((\llrr{p(t)}\sigma_0^2-\llrr{q(t)})\sigma_\tau^2\tau\right.\\
    \label{eq:appendix_hamiltonian_linearScaling_end}
    &\quad\left.-\llrr{q(t)}(\sigma_0^2-\sigma_\tau^2)t\right).
\end{align}
By solving Eqs.~\eqref{eq:appendix_hamiltonian_linearScaling_begin} to \eqref{eq:appendix_hamiltonian_linearScaling_end}, we can calculate $\expval{W}$ given by Eq.~\eqref{eq:dWdT_for_linear_scaling}.

Next, we derive $W_{\rm diss}$ of the HJE for the scaling protocol with time-dependent mass. With the setting in Section \ref{subsection:scaling}, by solving the Hamiltonian dynamics with an initial condition $(q_0,p_0)$, we have the final state $(q(\tau),p(\tau))$ as
\begin{align}
    q(\tau)&=e^{\frac{\gamma}{2}}(c_1\cos\omega\tau+c_2\sin\omega\tau),\\
    \nonumber
    p(\tau)&=\frac{1}{\sigma_0^2}e^{-\frac{\gamma}{2}}\Bigl\{\frac{\gamma}{2\tau}(c_1\cos\omega\tau+c_2\sin\omega\tau)\\&\quad+(-c_1\omega\sin\omega\tau+c_2\omega\cos\omega\tau)\Bigr\},
\end{align}
where
\begin{align}
    \omega&=\frac{\sqrt{4\tau^2/\alpha-\gamma^2}}{2\tau},\\
    c_1&=q_0,\\
    c_2&=\frac{1}{\omega}\qty(\frac{\sigma_0^2}{\alpha}p_0-\frac{\gamma}{2\tau}q_0)
\end{align}
for $\tau>\sqrt{\alpha}|\gamma|/2$. Then we have the distribution at $t=\tau$ and we obtain $W_{\rm diss}$ in Eq.~\eqref{eq:Wdiss_HJE_scaling_time_dependent_mass} by calculating the KL divergence in the same way as in Appendix \ref{appendix:parallel_Wdiss_HJE}.

\section{$W_{\rm diss}$ of the LJE for parallel transport and scaling}
\label{appendix:Wdiss_LJE_paralellVar}
We derive $W_{\rm diss}$ of the LJE for the protocol that includes both parallel transport and scaling. We obtain differential equations for the average and variance of position $\expval{q(t)}$ and $\llrr{q(t)}$ by substituting $U(q;t)$ in Section \ref{subsection:parallelVar} for Eqs.~\eqref{eq:appendix_langevin_mu} and \eqref{eq:appendix_langevin_var}:
\begin{align}
    \label{eq:appendix_langevin_mu_parallelVar}
    \dv{\expval{q(t)}}{t}&=\frac{a\mu t^2-(a\expval{q(t)}-\mu)\tau t-\tau^2\expval{q(t)}}{\tau^2 \sigma^2},\\
    \label{eq:appendix_langevin_var_parallelVar}
    \dv{\llrr{q(t)}}{t}&=-2\frac{a\llrr{q(t)} t+(\llrr{q(t)}-\sigma^2)\tau}{\tau\sigma^2}.
\end{align}
Then, the average work exerted during the process is
\begin{align}
    \nonumber
    \expval{W}&=\expval{\int_0^\tau\pdv{U(q;t)}{t}\dd t}\\
    \nonumber
    &=\int_0^\tau{\left[\frac{a}{2\sigma \tau}\left\{\llrr{q(t)}+\expval{q(t)}^2-\frac{2\mu t}{\tau}\expval{q(t)}+\qty(\frac{\mu t}{\tau})^2\right\}\right.}\\
    \label{eq:appendix_langevin_work}
    &\quad-\frac{1+a\frac{t}{\tau}}{\sigma^2}\frac{\mu}{\tau}\qty(\expval{q(t)}-\frac{\mu t}{\tau})\Biggr]  \dd t.
\end{align}
We can calculate Eq.~\eqref{eq:appendix_langevin_work} by solving Eqs.~\eqref{eq:appendix_langevin_mu_parallelVar} and \eqref{eq:appendix_langevin_var_parallelVar} numerically.
Finally, we obtain $W_{\rm diss}$ by subtracting $\Delta F=1/2\log(1+a)$ from $\expval{W}$.

\section{$W_{\rm diss}$ of the HJE for parallel transport and scaling}
\label{appendix:Wdiss_HJE_paralellVar}
We derive $W_{\rm diss}$ of the HJE for the protocol that includes both parallel transport and scaling. In the same manner as in Appendix \ref{appendix:Wdiss_HJE_scaling}, we obtain the time derivatives of states' average $\langle\,\cdot\,\rangle$, variance $\llrr{\,\cdot\,}$, and covariance $\llrr{\cdot,\cdot}$ at time $t$ with the setting in Section \ref{subsection:parallelVar}:
\begin{align}
    \label{eq:appendix_hamiltonian_parallelVar_begin}
    \dv{\expval{q(t)}}{t}&=\expval{p(t)},\\
    \dv{\expval{p(t)}}{t}&=\frac{a\mu t^2-(a\expval{q(t)}-\mu)t\tau-\tau^2\expval{q(t)}}{\tau^2\sigma^2},\\
    \dv{\llrr{q(t)}}{t}&=2\llrr{q(t), p(t)},\\
    \dv{\llrr{p(t)}}{t}&=-\frac{2\llrr{q(t), p(t)}(at+\tau)}{\tau\sigma^2},\\
    \label{eq:appendix_hamiltonian_parallelVar_end}
    \dv{\llrr{q(t), p(t)}}{t}&=-\frac{at\llrr{q(t)}-(\sigma^2 \llrr{p(t)}-\llrr{q(t)})\tau}{\tau\sigma^2}.
\end{align}
Then, the average work $\expval{W}$ is described by Eq.~\eqref{eq:appendix_langevin_work}, which is the same as that used for the LJE. This is because the work exerted during a process only depends on a potential energy function.

We can calculate $\expval{W}$ by solving Eqs.~\eqref{eq:appendix_hamiltonian_parallelVar_begin} to \eqref{eq:appendix_hamiltonian_parallelVar_end} numerically and finally obtain $W_{\rm diss}$ by subtracting $\Delta F=1/2\log(1+a)$ from $\expval{W}$.

\section{$W_{\rm diss}$ of the LJE for rotation}
\label{appendix:Wdiss_LJE_rotation}
We derive $W_{\rm diss}$ of the LJE for the rotation protocol. In the same manner as in Appendix \ref{appendix:Wdiss_LJE_scaling}, we obtain the time derivatives of states' average $\langle\,\cdot\,\rangle$, variance $\llrr{\,\cdot\,}$, and covariance $\llrr{\cdot,\cdot}$ at time $t$ with the setting in Section \ref{section:Rotation}:
\begin{align}
\nonumber
    \nonumber
    \dv{\llrr{x(t)}}{t}&=-4\llrr{x(t)}\qty(a\cos\theta(t)^2+b\sin\theta(t)^2)\\
    \label{eq:LJE_rotation_exp_begin}
    &\quad+4\llrr{x(t),y(t)}(a-b)\cos\theta(t)\sin\theta(t)+2,\\
    \nonumber
    \dv{\llrr{y(t)}}{t}&=-4\llrr{y(t)}\qty(b\cos\theta(t)^2+a\sin\theta(t)^2)\\
    &\quad+4\llrr{x(t),y(t)}(a-b)\cos\theta(t)\sin\theta(t)+2,\\
    \nonumber
    \dv{\llrr{x(t),y(t)}}{t}&=2\qty(\llrr{x(t)}+\llrr{y(t)})(a-b)\times\\
    \label{eq:LJE_rotation_exp_end}
    &\quad\cos\theta(t)\sin\theta(t)-2\llrr{x(t),y(t)}(a+b).
\end{align}

Then, the time derivative of $U(x,y;t)$ given by Eq.~\eqref{eq:rotation_U)} is
\begin{align}
    \label{eq:rotation_dudt}
    \begin{split}
        \pdv{U(x,y;t)}{t}&=2\dv{\theta}{t}\{(a-b)\sin\theta(t)\cos\theta(t)(-x^2+y^2)\\&\quad-(a-b)(\cos^2\theta-\sin^2\theta)xy \}.
    \end{split}
\end{align}
We describe $W_{\rm diss}$ by $x$, $y$ and $\theta$ with the expected value of Eq.~\eqref{eq:rotation_dudt}:
\begin{align}
    W_{\rm diss}
    \nonumber
    &=\expval{\int_0^\tau \pdv{U(x,y;t)}{t}\dd t}\\
    \nonumber
    &=\int_0^\tau\expval{\pdv{U(x,y;t)}{t}}\dd t\\
    \label{eq:rotation_Wdiss_LJE_int}
    \begin{split}
        &=\int_0^\tau 2\dv{\theta}{t}\{(a-b)\sin\theta(t)\cos\theta(t)(-\expval{x^2}+\expval{y^2})\\&\quad-(a-b)(\cos^2\theta(t)-\sin^2\theta(t))\expval{xy}\dd t \}.
    \end{split}
\end{align}
The expected values in Eq.~\eqref{eq:rotation_Wdiss_LJE_int} can be obtained by numerically solving Eqs.~\eqref{eq:LJE_rotation_exp_begin} to \eqref{eq:LJE_rotation_exp_end}.

\section{$W_{\rm diss}$ of the HJE for rotation}
\label{appendix:Wdiss_HJE_rotation}
We derive $W_{\rm diss}$ of the HJE for the rotation protocol. In the same manner as in Appendix \ref{appendix:Wdiss_HJE_scaling}, we obtain the time derivatives of states' average $\langle\,\cdot\,\rangle$ at time t under the setting in Section \ref{section:Rotation}:
\begin{align}
    \label{eq:rotation_exp_1}
    \frac{d\langle x^2\rangle}{dt}&=2\langle xp_x\rangle,\\
    \frac{d\langle y^2\rangle}{dt}&=2\langle yp_y\rangle,\\
    \nonumber
    \frac{d\langle p_x^2\rangle}{dt}&=-4(a\cos^2\theta(t)+b\sin^2\theta(t))\langle xp_x\rangle\\&\quad+4(a-b)\sin\theta(t)\cos\theta(t)\langle yp_x\rangle,\quad\quad\quad
\end{align}
\begin{align}
\nonumber
\frac{d\langle p_y^2\rangle}{dt}&=-4(b\cos^2\theta(t)+a\sin^2\theta(t))\langle yp_y\rangle\\&\quad+4(a-b)\sin\theta(t)\cos\theta(t)\langle xp_y\rangle,\\
\nonumber
  \frac{d\langle xp_x\rangle}{dt}&=-2(a\cos^2\theta(t)+b\sin^2\theta(t))\langle x^2\rangle\\&\quad+2(a-b)\cos\theta(t)\sin\theta(t)\langle xy\rangle+\langle p_x^2\rangle,\\
  \frac{d\langle xy\rangle}{dt}&=\langle xp_y\rangle +\langle yp_y\rangle ,\\
 \nonumber
  \frac{d\langle xp_y\rangle}{dt}&=-2(b\cos^2\theta(t)+a\sin^2\theta(t))\langle xy\rangle\\&\quad +2(a-b)\sin\theta(t)\cos\theta(t)\langle x^2\rangle+\langle p_xp_y\rangle,\\
 \nonumber
  \frac{d\langle yp_x\rangle}{dt}&=-2(a\cos^2\theta(t)+b\sin^2\theta(t))\langle xy\rangle\\&\quad +2(a-b)\sin\theta(t)\cos\theta(t) \langle y^2\rangle+\langle p_xp_y\rangle,\\
 \nonumber
  \frac{d\langle p_x p_y\rangle}{dt}&=-2(a\langle xp_y\rangle +b\langle yp_y\rangle)\cos^2\theta(t)\\&\quad +2(a-b)\sin\theta(t)\cos\theta(t)(\langle xp_x\rangle +\langle yp_y\rangle)\nonumber\\&\quad-2(a\langle yp_x\rangle+b\langle xp_y\rangle)\sin^2\theta(t),\\
  \label{eq:rotation_exp_10}
 \nonumber
  \frac{d\langle yp_y\rangle}{dt}&=-2(b\cos^2\theta(t)+a\sin^2\theta(t))\langle y^2\rangle\\&\quad+2(a-b)\sin\theta(t)\cos\theta(t)\langle xy\rangle +\langle p_y^2\rangle
\end{align}

The expected values $\expval{W}$ given by Eq.~\eqref{eq:rotation_Wdiss_LJE_int} can be obtained by numerically solving Eqs.~\eqref{eq:rotation_exp_1} to \eqref{eq:rotation_exp_10}.

\section{$Z$ for a multimodal distribution}
\label{appendix:Z_of_multimodal}
We calculate the partition function $Z$ of the target distribution given by Eq.~\eqref{eq:coshManyPeaks_f} in Section \ref{section:multimodal}. Considering the superposition of distributions, $Z$ is given by
\begin{align}
    \label{eq:Z_of_multimodal_pre}
    Z=N\int_{-\infty}^\infty{e^{-\cosh{\|q\|_2}}}\dd q.
\end{align}
Therefore, we focus on the integration in the right-hand side. Let $S_{D-1}$ be the surface area of $D-1$ dimensional unit sphere:
\begin{align}
    S_{D-1}=\frac{2\pi^{D/2}}{\Gamma(D/2)},
\end{align}
where $\Gamma(x)$ is the Gamma function. In $D$-dimensional polar coordinates, Eq.~\eqref{eq:Z_of_multimodal_pre} is
\begin{align}
    \label{eq:Z_of_multimodal_last}
    Z =  NS_{D-1}\int_0^\infty{e^{-\cosh{r}}r^{D-1}}\dd r.
\end{align}
We can obtain $Z$ for $D=16$ and arbitrary $N$ by numerically calculating Eq.~\eqref{eq:Z_of_multimodal_last}.

\end{document}